\documentclass[trackchanges,twocolumn]{aastex701}

\begin{document}

\title{Warped Disk Galaxies: \\ Alignment with the Large-Scale Tidal Field}

\author[0009-0004-6972-0193]{Yiheng Wang}
\email{yihengwang@pmo.ac.cn}
\affiliation{Purple Mountain Observatory, Chinese Academy of Sciences, Nanjing 210023, Jiangsu, China}
\affiliation{School of Astronomy and Space Sciences, University of Science and Technology of China, Hefei, Anhui 230026, China}

\author[0000-0002-4911-6990]{Huiyuan Wang}
\email{whywang@ustc.edu.cn}
\affiliation{Department of Astronomy, University of Science and Technology of China, Hefei, 230026, China}
\affiliation{School of Astronomy and Space Sciences, University of Science and Technology of China, Hefei, Anhui 230026, China}

\author[0000-0003-1588-9394]{Enci Wang}
\email{12345}
\affiliation{Department of Astronomy, University of Science and Technology of China, Hefei, 230026, China}
\affiliation{School of Astronomy and Space Sciences, University of Science and Technology of China, Hefei, Anhui 230026, China}

\author[0000-0002-5458-4254]{Xi Kang}
\email{kangxi@zju.edu.cn}
\affiliation{Institute for Astronomy, School of Physics, Zhejiang University, Hangzhou 310058, Zhejiang, China}
\affiliation{Center for Cosmology and Computational Astrophysics, Zhejiang University, Hangzhou 310058, Zhejiang, China}
\affiliation{Purple Mountain Observatory, Chinese Academy of Sciences, Nanjing 210023, Jiangsu, China}

\correspondingauthor{Xi Kang}
\email{kangxi@zju.edu.cn}
 
\begin{abstract}

A possible origin of disk galaxy warps is the misalignment between galactic disks and their host dark matter halos, the orientations of which are found to be statistically aligned with the large-scale tidal field. In this work, we test this scenario by examining the alignment between warped disk galaxies and the large-scale tidal field reconstructed from the ELUCID project. We find a statistically significant alignment signal between disk orientations and the $t_1$ direction, with warped and non-warped galaxies showing different alignment behaviors. Warped galaxies show an excess of intermediate angles and a preference for orientations slightly offset from perfect parallel and perpendicular alignments. In contrast, non-warped galaxies exhibit a deficit of intermediate angles relative to random expectations, which becomes more pronounced after matching to a control sample. We also find a clear mass dependence, with high-mass warped galaxies contributing the excess of intermediate-angle signal. No significant alignment signal in warped galaxies is detected with the $t_3$ direction.

\end{abstract}


\section{Intrduction} \label{sec:intro}

The orientations of galaxies encode preferred alignments shaped by the large-scale structure of the Universe and their host dark matter halos \citep[e.g.,][]{Kiessling_2015, Kirk_2015, Joachimi_2015}. In the hierarchical framework of structure formation, galaxies form and evolve within dark matter halos that assemble through gravitational clustering and grow via anisotropic accretion within the cosmic web \citep[e.g.,][]{White_1978, Mo_2010, Bond_1996, Springel_2005, Dekel_2009}. The large-scale tidal field of the cosmic web induces anisotropic collapse and tidal torques, thereby shaping the orientations and angular momenta of dark matter halos and their embedded galaxies \citep[e.g.,][]{Catelan_2001, Hirata_2004, Dubois_2014, Codis_2015}. Consequently, galaxies tend to exhibit preferred orientations with respect to their host dark matter halos, neighboring galaxies, and the surrounding large-scale structure \citep[e.g.,][]{Jing_2002, Mandelbaum_2006, Yang_2006, Hirata_2007, Faltenbacher_2009, Wang_2018, Xu_2023}.

In dark matter simulations, halo shapes are aligned with the eigenvectors of the large-scale tidal field, with the halo major and minor axes aligning with the slowest- and fastest-collapse directions, respectively\citep[e.g.,][]{Wang_2011, Libeskind_2013, Chen_2016}. Halo spins, on the other hand, preferentially align with the intermediate axis. Baryonic effects and galaxy–halo misalignments partially weaken the direct correspondence between galaxies and the underlying dark matter structure\citep[e.g.,][]{Dubois_2014, Codis_2015, Wang_2018, Zhang_2023}. Observational studies have found that galaxy alignments with the large-scale tidal field depend strongly on galaxy type. Massive, red galaxies exhibit measurable shape alignments, with their stellar major axes preferentially aligning with the slowest collapsing direction of the large-scale structure \citep[e.g.,][]{Zhang_2013}, while spiral galaxies typically exhibit only weak or marginal alignments between their spin axes and the tidal field\citep[e.g.,][]{Zhang_2015}. 

One proposed mechanism for the formation of galactic warps is the misalignment between the disk and its host dark matter halo\citep[e.g.,][]{binney_1992, dubinski_1995, Debattista_1999}. When there is a misalignment between the disk and the halo, tidal forces can induce torques that lead to the warping of the galaxy's disk\citep{Han_2023}. Such torques are strongest at intermediate misalignment angles, but vanish when the disk and halo are either perfectly aligned or mutually orthogonal\citep{wang_2026}. Since dark matter halos are found to be aligned with the large-scale tidal field, this scenario naturally predicts that warped disk galaxies may exhibit distinct alignment signatures with the large-scale tidal field.

The anisotropic distribution of satellite galaxies has also been used as a statistical tracer of halo orientation. By stacking central galaxies with multiple satellites, \cite{Zee_2025} found that the satellite distribution around warped disk galaxies exhibits an excess around 45°, whereas satellites around non-warped disk galaxies are consistent with an approximately random distribution. This stacking method is limited to systems with a sufficiently large number of satellite galaxies. To expand the study of disk-halo misalignment across a broader population, we use tidal field directions as tracers of halo orientations in this work, examining their alignment with warped and non-warped disk galaxies.

This paper is organized as follows. In Section~2, we describe the catalog of warped disk galaxies and the large-scale tidal field data used in our analysis. Section~3 presents our main results, and Section~4 summarizes our conclusions. The cosmological parameters adopted in this work are consistent with those used in the ELUCID project, corresponding to a WMAP5 cosmology with $\Omega_m = 0.258$, $\Omega_\Lambda = 0.742$, and $H_0 = 72 \, \mathrm{km \, s^{-1} \, Mpc^{-1}}$.

\section{Data and Methodology} \label{sec:meth}
\subsection{Warped Disk Galaxies}
In our previous work, we used a deep-learning approach to classify edge-on disk galaxies in the Galaxy Zoo DESI \citep{Walmsley_2023} into warped and non-warped categories, and constructed high-confidence catalogs for both. In this paper, we select the subsample within the ELUCID volume \citep{Wang_2016}, the \textit{Exploring the Local Universe with the Reconstructed Initial Density Field} project, for which a reconstructed large-scale tidal field is available.

After crossed match our high-confidence catalogs with reconstruted samples in ELUCID using a 2 arcsecond radius, we obtained 2,053 warped samples and 18,050 non-warped samples. We matched dr8\_id with the external catalog released by \cite{Walmsley_2023} to obtain other properties of these galaxies, including stellar mass, color, redshift and position angles from SDSS DR7 measurements\citep[e.g.,][]{Abazajian_2009, yang_2007}. The disk orientation is determined by the position angle (PA), which is defined as the angle measured east of north, ranging from 0° to 180°. We adopt the Sérsic model position angle, $sersic\_phi$, rather than the Petrosian-based $elpetro\_phi$, since the former provides a more stable estimate of the global disk orientation and is less sensitive to outer asymmetries and warp-induced distortions.

\begin{figure*}[htbp]
  \centering
  \includegraphics[width=\textwidth]{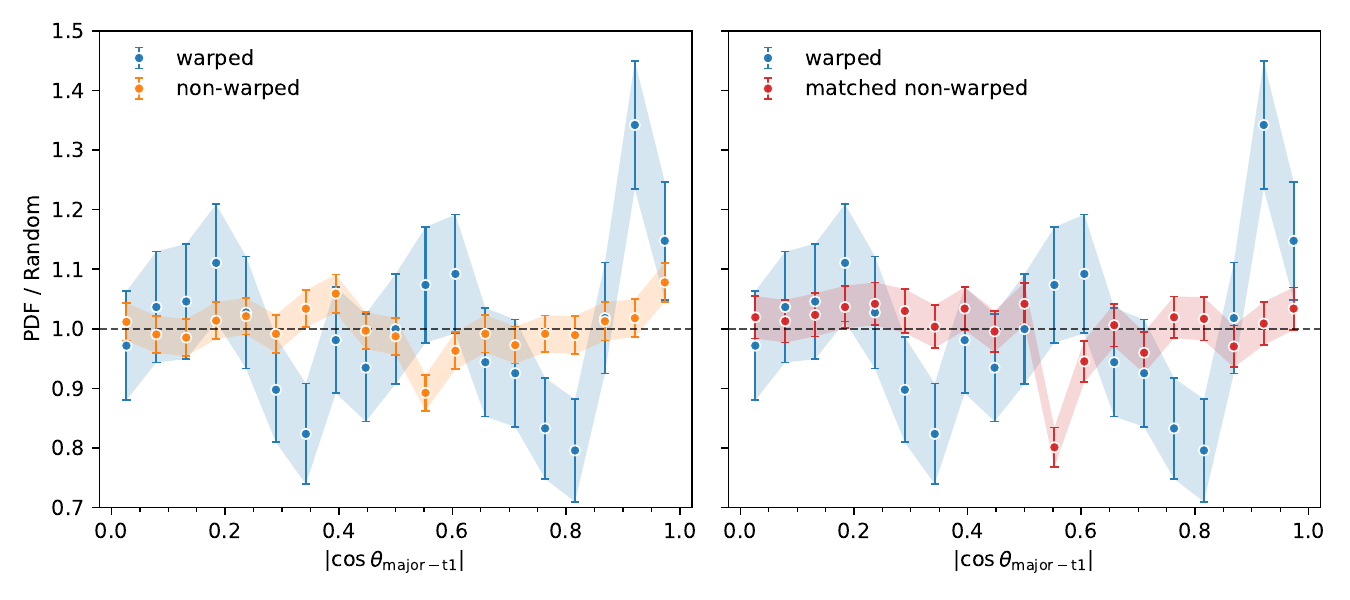}
  \caption{Distribution of $|\cos\theta|$ between the galaxy major axis and the tidal field $t_1$ direction for warped galaxies (blue), non-warped galaxies (orange), and a matched non-warp control sample (red). The curves show probability density functions (PDFs) computed from the data and normalized to a uniform distribution. The shaded regions represent $1\sigma$ uncertainties estimated via bootstrap resampling ($N_{\rm boot}=2000$). The dashed horizontal line shows the uniform distribution expectation. The left panel compares warped galaxies to the full non-warped sample, and the right panel compares warped galaxies to the matched control sample.}
  \label{fig:major_t1}
\end{figure*}

Given a galaxy at equatorial coordinates $(\alpha, \delta)$, the unit line-of-sight vector in the global Cartesian coordinate system is
\begin{equation}
\hat{\mathbf{r}} =
(\cos\delta \cos\alpha,\,
 \cos\delta \sin\alpha,\,
 \sin\delta).
\end{equation}

We define the local tangent-plane basis vectors as
\begin{equation}
\hat{\mathbf{e}}_E =
(-\sin\alpha,\,
 \cos\alpha,\,
 0),
\end{equation}

\begin{equation}
\hat{\mathbf{e}}_N =
(-\sin\delta \cos\alpha,\,
 -\sin\delta \sin\alpha,\,
 \cos\delta).
\end{equation}

The position angle $\phi$ defines the orientation of the projected major axis in the tangent plane, and the corresponding unit vector in the local frame is
\begin{equation}
\hat{\mathbf{n}}_{\mathrm{local}} =
\cos\phi \, \hat{\mathbf{e}}_N +
\sin\phi \, \hat{\mathbf{e}}_E.
\end{equation}

The corresponding unit vector in the global Cartesian coordinate system is obtained via a basis transformation:
\begin{equation}
n_x =
-\cos\phi\,\sin\delta\cos\alpha
-\sin\phi\,\sin\alpha
\end{equation}

\begin{equation}
n_y =
-\cos\phi\,\sin\delta\sin\alpha
+\sin\phi\,\cos\alpha
\end{equation}

\begin{equation}
n_z =
\cos\phi\,\cos\delta
\end{equation}

\subsection{Large-scale Tidal Field}

In this work, we adopt the gravitational potential field provided by the ELUCID project, which reconstructs the mass density field based on galaxy groups identified\citep{yang_2007} in the SDSS survey\citep{Blanton_2005, Abazajian_2009}. The tidal tensor is defined as the Hessian Matrix of the gravitational potential:

\begin{equation}
T_{ij} = \frac{\partial^2 \Phi}{\partial x_i \partial x_j}, 
\quad i,j = x,y,z,
\end{equation}

where $\Phi$ is the gravitational potential. Diagonalizing the tidal tensor yields three eigenvalues 
($\lambda_1 \ge \lambda_2 \ge \lambda_3$) and corresponding eigenvectors ($\mathbf{t}_1, \mathbf{t}_2, \mathbf{t}_3$). 
These eigenvectors represent the principal directions of the large-scale tidal field, where $\mathbf{t}_1$ corresponds to the direction of fastest collapse and is expected to align with the minor (short) axis of dark matter halos, $\mathbf{t}_3$ corresponds to the slowest collapse direction and is typically associated with the major axis, and $\mathbf{t}_2$ denotes the intermediate axis.

For each galaxy, we estimate the local tidal field at its position using trilinear interpolation from the eight nearest grid points of the reconstructed tidal field, thereby obtaining the tidal eigenvectors at the galaxy's location. We also compute the tidal directions at the nearest grid point as an alternative estimate. Due to the smoothing scale adopted in the ELUCID reconstruction, the resulting alignment signals from the two methods are found to be nearly identical.

Previous studies of disk galaxy alignments with the large-scale tidal field have primarily focused on the alignment between galaxy spin vectors and the tidal field. The spin direction of spiral galaxies is often inferred from observed axis ratios and position angles. In edge-on systems, uncertainties in the intrinsic thickness and projection effects can introduce errors in the reconstructed spin vectors. In this work, our primary goal is to study the geometric alignment between disk orientations and the large-scale tidal field, particularly the $\mathbf{t}_1$ and $\mathbf{t}_3$ directions, which correspond to the directions of strongest tidal compression and slowest collapse and are expected to trace the minor and major-axis directions of dark matter halos, respectively. We therefore use the disk position angle (PA) as a robust tracer of the projected disk orientation, and measure its alignment with both the $\mathbf{t}_1$ and $\mathbf{t}_3$ axes of the tidal field, rather than attempting to reconstruct the spin vector.

\section{Results} \label{sec:result}

\subsection{Alignment with Tidal Fields}
Our results are shown in Figure~\ref{fig:major_t1}. We find that warped and non-warped galaxies exhibit distinctly different alignment behaviors in the angle between the disk direction and the large-scale $t_1$ direction.

For warped galaxies, the distribution shows an excess relative to a random distribution at intermediate angles, as well as at angles slightly offset from the perfectly perpendicular and perfectly parallel configurations (around $20^\circ$ and $80^\circ$). In contrast, the full non-warped sample (which contains a larger number of galaxies) is broadly consistent with a random distribution, except for a deficit at intermediate angles. Within the framework of the misaligned halo model, the tidal torque exerted by a misaligned dark matter halo vanishes only when the disk and halo are perfectly parallel or perpendicular, and increases toward intermediate misalignment angles, reaching a maximum near $45^\circ$. The excess of warped galaxies at intermediate angles is therefore qualitatively consistent with the expectation that stronger halo torques facilitate the formation of disk warps. The additional excesses around $20^\circ$ and $80^\circ$ further suggest that even the weaker torques associated with modest departures from the perfectly parallel or perpendicular configurations may be sufficient to induce warps. Conversely, the deficit of non-warped galaxies at intermediate angles suggests that this angular range is statistically disfavored for sustaining non-warped configurations.

To minimize the influence of confounding factors, we further perform a matched-control analysis for the non-warped sample, adopting the same matching strategy as in our previous work\citep{Wang_2025}. For each warped galaxy, we identify seven closest non-warped counterparts in the three-dimensional parameter space of stellar mass, color, and redshift. After the matching procedure, the deficit at intermediate angles in the matched non-warped sample becomes more pronounced, falling to about $20\%$ below the random expectation and indicating a statistically significant alignment signal. This result further supports the interpretation that galaxies with intermediate halo--disk misalignment angles are less stable in maintaining non-warped configurations and are therefore more likely to develop disk warps.

\begin{figure*}
    \centering
    \includegraphics[width=0.9\textwidth,height=0.9\textheight,keepaspectratio]{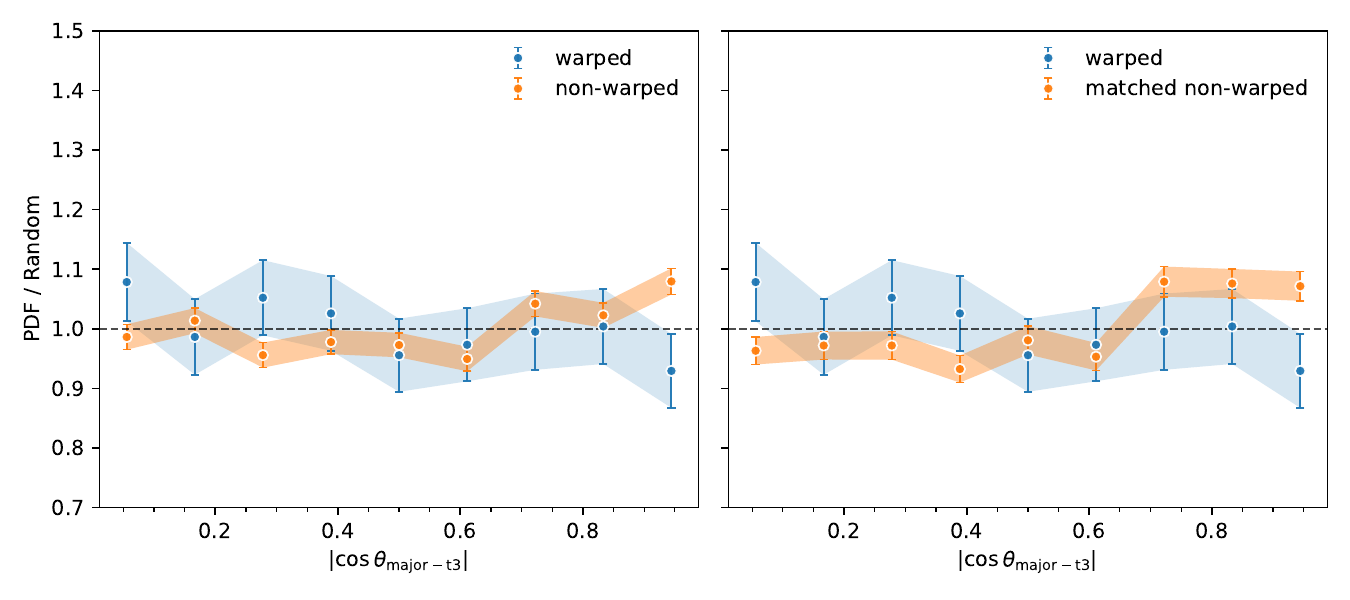}
    \caption{Alignment between galaxy major axis and the large-scale tidal field direction $t_3$, quantified using $|\cos\theta|$. The plotting conventions are the same as in Fig.~\ref{fig:major_t1} for $t_1$.}
    \label{fig:major_t3}
\end{figure*}

Non-warped galaxies show a weak tendency toward alignment with the $t_3$ direction, while warped galaxies are consistent with no detectable alignment signal (Figure~\ref{fig:major_t3}). In numerical simulations, the $t_3$ direction of the tidal field is known to be aligned with the major axis of dark matter halos and can also serve as a tracer of the filamentary structure of the cosmic web. The absence of a detectable signal may reflect both physical and methodological effects. On the physical side, disk configurations are expected to be more stable when the disk plane is perpendicular to the halo minor-axis ($t_1$) direction than to the halo major-axis ($t_3$) direction, implying that deviations from this equilibrium configuration are more likely to produce a measurable alignment signal for warped galaxies with respect to $t_1$. Also, the $t_1$ direction, corresponding to the largest eigenvalue of the tidal tensor, is generally more robustly reconstructed than the $t_3$ direction, which is more sensitive to noise and eigenvalue degeneracies. 

\subsection{Dependence on Galaxy Masses}

\begin{figure*}
  \centering
  \includegraphics[width=0.9\textwidth,height=0.9\textheight,keepaspectratio]{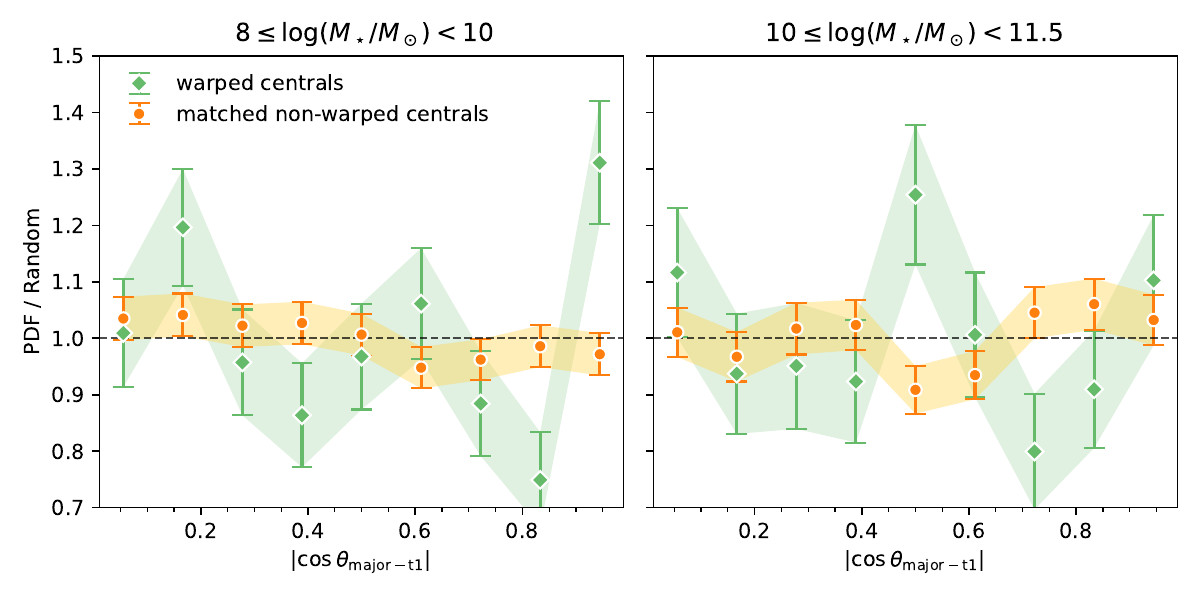}
  \caption{Distribution of $|\cos\theta|$ between the galaxy major axis and the tidal field $t_1$ direction for central galaxies in two stellar mass bins, using warped galaxies and a matched non-warped control sample. The left and right panels correspond to $8 \leq \log(M_\star/M_\odot) < 10$ and $10 \leq \log(M_\star/M_\odot) < 11.5$, respectively. The curves show probability density functions (PDFs) computed from the data normalized relative to a uniform distribution. The shaded regions represent $1\sigma$ uncertainties estimated via bootstrap resampling ($N_{\rm boot}=2000$).}
  \label{fig:mass_bin}
\end{figure*} 

We further divide the sample into stellar-mass bins. To ensure sufficient statistics for the alignment measurements, we split the galaxies at $10^{10}M_\odot$, resulting in two bins, $8 \le \log(M_*/M_\odot) < 10$ and $10 \le \log(M_*/M_\odot) < 11.5$, with comparable sample sizes in each bin. Stellar masses are taken from the NASA-Sloan Atlas (NSA) catalog using the parameter \texttt{elpetro\_mass\_log} \citep{Blanton_2011}, which is derived from \texttt{kcorrect}-based template fitting to the SDSS ugriz photometry and assumes a Chabrier initial mass function \citep{Chabrier_2003}. 

We find that low-mass warped galaxies mainly contribute to the excess at small angles offset from the perfectly perpendicular and perfectly parallel configurations (i.e., the peaks around $\sim 20^\circ$ and $\sim 80^\circ$). In contrast, high-mass warped galaxies primarily contribute to the excess at intermediate angles (Figure~\ref{fig:mass_bin}). 

These mass-dependent trends are naturally explained within the framework of the misaligned halo model. Lower-mass disk galaxies appear to develop warps even when the halo--disk misalignment is only slightly offset from the perfectly parallel or perpendicular configurations, suggesting that relatively modest tidal torques may already be sufficient to warp their disks. By contrast, more massive disk galaxies require stronger tidal torques to be warped, and the tidal torque amplitude peaks at intermediate halo--disk misalignment angles, consistent with a $\sin(2\theta)$ dependence.

\section{Summary} \label{sec:sum}

In our previous work, we found that a substantial fraction of warped galaxies lack any nearby companions within $\sim 100\,\mathrm{kpc}$, suggesting that galaxy–galaxy interactions alone cannot account for warp formation in many cases. Motivated by this result, we investigate whether a misalignment between dark matter halos and galactic disks leaves observable signatures in the alignment signals of warped and non-warped galaxies.

To this end, we study the alignment between disk orientations of warped galaxies and the large-scale tidal field reconstructed in the ELUCID project. The tidal tensor is adopted as a proxy for the underlying dark matter halo orientation. In particular, we focus on the projected disk direction of edge-on galaxies and examine its relation to the eigen-directions of the tidal field, $t_1$ and $t_3$, where $t_1$ typically aligns with the minor axis of dark matter halos, while $t_3$ corresponds to the halo major axis and is closely associated with the filamentary structure of the cosmic web.

Our main results are summarized as follows:

1. Warped and non-warped disk galaxies exhibit clearly different alignment behaviors. Warped galaxies show significant excess signals relative to random at intermediate angles, as well as at angles slightly offset from the perfectly parallel and perpendicular configurations (around $\sim 20^\circ$ and $\sim 80^\circ$). In contrast, the full non-warped sample is much closer to a random distribution, except for a deficit at intermediate angles.

2. After performing a matched-control analysis to remove the influence of stellar mass, color, and redshift, the control non-warped sample shows a more pronounced deficit at intermediate angles, with a suppression of about $20\%$ relative to random. The result is consistent with a scenario in which the halo-induced tidal torque is strongest at intermediate misalignment angles, making it difficult for disk galaxies to remain in a non-warped configuration.

3. The alignment signal depends on stellar mass. Low-mass warped galaxies mainly contribute to the excess near $\sim 20^\circ$ and $\sim 80^\circ$, while high-mass warped galaxies predominantly drive the excess at intermediate angles. This behavior is consistent with a physical picture in which tidal torques from misaligned halos peak at intermediate misalignment angles, and more massive disks require larger torques to develop observable warps.

4. No significant alignment signal in warped galaxies is detected with the $t_3$ direction of the tidal field, which is associated with the halo major axis and the surrounding filamentary structure. In contrast, non-warped galaxies show a weak tendency for their galactic major axes to align parallel to the $t_3$ direction.

The distinct alignment patterns between warped and non-warped galaxies shown above, together with the observed deficit at intermediate angles in non-warped systems and the mass dependence of the alignment signal, are consistent with a disk–halo misalignment scenario. Within this framework, warped disk galaxies may serve as probes of halo–disk misalignment, highlighting the potential role of tidal torques in shaping the outer structure of disk galaxies.

\begin{acknowledgments}
This work is supported by the National Key Research and Development Program (No. 2022YFA1602903, No. 2023YFB3002502), NSFC(No. 12595314, No. 12533007, No. 12547104), the science research grants from the China Manned Space project with NO.CMS-CSST-2025-A10. HYW is supported by the National Natural Science Foundation of China (NSFC, Nos. 12192224). EW thanks support of the National Science Foundation of China (Nos. 12473008). The model training and data analysis were performed on the SilkRiver high-performance computing platform at Zhejiang University.
\end{acknowledgments}

\bibliography{ref}{}
\bibliographystyle{aasjournalv7}

\end{document}